\documentclass[twoside]{ae100prg}
\usepackage{graphicx}
\usepackage[breaklinks]{hyperref}
\usepackage{booktabs}

\usepackage{ifpdf,epsfig,array,amsthm,amssymb,psfrag}
\usepackage[T1]{fontenc}
\usepackage{aeguill}

\DeclareFontFamily{OT1}{rsfs}{} \DeclareFontShape{OT1}{rsfs}{m}{n}{
<-7> rsfs5 <7-10> rsfs7 <10-> rsfs10}{}
\DeclareMathAlphabet{\mathscr}{OT1}{rsfs}{m}{n}

\DeclareFontFamily{OT1}{rsfs}{} \DeclareFontShape{OT1}{rsfs}{m}{n}{
<-7> rsfs5 <7-10> rsfs7 <10-> rsfs10}{}
\DeclareMathAlphabet{\mycal}{OT1}{rsfs}{m}{n}

\newcommand{\R}{\mathbb{R}}
\newcommand{\C}{\mathbb{C}}
\newcommand{\I}{\mathrm{i}}

\begin{document}
\title{Superradiance or total reflection?}


\author{Andr\'as L\'aszl\'o$^{1,2}$\, Istv\'an R\'acz$^2$}

\address{$^1$ CERN, CH-1211 Gen\'eve 23, Switzerland}
\address{$^2$ Wigner RCP, H-1121 Budapest,
Hungary}

\email{laszlo.andras@wigner.mta.hu, racz.istvan@wigner.mta.hu}

\begin{abstract}
Numerical evolution of massless scalar fields on Kerr background is studied. 
The initial data specifications are chosen to have compact support separated from the  
ergoregion and to yield nearly monochromatic incident wave packets. The initial data 
is also tuned to maximize the effect of superradiance. Evidences are shown indicating 
that instead of the anticipated energy extraction from black hole the incident radiation 
fail to reach the ergoregion rather it suffers a nearly perfect reflection.
\end{abstract}

\section{Introduction}

To motivate our investigations let us mention first that the stability of the Kerr family of 
black hole solutions within the space of the vacuum solutions to the Einstein equations is one 
of the most important unresolved issues in general relativity. The ultimate goal is 
to provide boundedness and decay statements for solutions of the vacuum Einstein equations around 
the members of the Kerr family. 

It may be a surprise that---even nowadays when numerical simulations of binary black hole systems 
becomes a daily routine---essentially all work concerning the aforementioned black hole stability 
problem has been confined to the linearized setting. Indeed, considerations are restricted to 
study the solutions to the  Klein-Gordon equation 
\begin{equation}\label{KG} 
\Box_K \Phi = 0\, 
\end{equation}  
on Kerr background. This is done with the hope that the understanding of these simplified scalar perturbations are good 
preparations to the study of more complicated problem of full but yet linear gravitational perturbations. 

It should also be mentioned that all the available analytic proofs justifying the linear stability 
for scalar perturbations, even in this simplest possible setting, are known to be restricted to the case of 
slowly rotating subextremal Kerr black holes \cite{LB,DR}.  

\section{Superradiance}

To start off recall first that superradiance was discovered in the early 70's as a new phenomenon 
and it may be associated with the names of Misner, Zel'dovich and Starobinskii \cite{misner,zeld,staro}. 
It is also considered to be the wave analog of the Penrose process and {\it it is supposed to allow energy 
to be extracted from black holes}. 

The common belief related to the interaction of black holes with incident radiation is 
summarized as ``...if scalar, electromagnetic or gravitational wave is incident upon a black hole,  
part of the wave ({the ``transmitted wave''}) will be absorbed by the black hole and part of the wave 
({ the ``reflected wave''}) will escape to infinity'' \cite{wald}. Recall that by using the 
Teukolsky equation \cite{teukolsky} the evolution of scalar, electromagnetic and gravitational 
perturbations can be investigated within 
the same setting. It is also important to be mentioned that all the conventional arguments ending up 
with superradiance, including the ones based on Teukolsky's equation, refer to properties of individual 
modes \cite{pressteukolsky}. 

Interestingly, as it was pointed out first by Bekenstein \cite{bekenstein}, whenever superradiance happens 
it can be seen to be completely consistent with the laws of black hole thermodynamics. It is also worth to be 
mentioned some of the expectations relevant for scalar perturbations. It was claimed in \cite{yau} ``Starobinskii made an 
asymptotic expansion for the reflection coefficient and found a relative gain of energy of about {${5\%}$} 
for $m = 1$ and less than {${1\%}$} for $m \geq 2$''. 

\section{Superradiance in mode analysis} 

It was realized first by Carter \cite{carter} that the temporal Fourier transform, 
${\mycal{F}}\hskip-0.5mm\Phi =\frac{1}{\sqrt{2\pi}}\int_{-\infty}^{+\infty}\Phi e^{i\omega t} dt\,,$
of a solution to (\ref{KG}) may be decomposed as 
\begin{equation}\label{kerrrepresentation}
{\mycal{F}}\hskip-0.5mm\Phi(\omega,r_{*},\vartheta,{\varphi})=\frac{1}{\sqrt{r^{2}+a^{2}}}
\sum_{\ell=0}^{\infty}\sum_{m=-\ell}^{\ell}R_{\ell,\omega}^{m}(r_{*})S_{\ell,a\omega}^{m}
(\vartheta ,{\varphi})\,,  
\end{equation}
where $t,r_{*},\vartheta,{\varphi}$ are local coordinates, while $\omega$ is the frequency in the time 
translation direction. In (\ref{kerrrepresentation}) 
$S_{\ell,a\omega}^{m}$ denotes the {\it oblate spheroidal harmonic functions}, with oblateness parameter 
$a\omega$, and with angular momentum quantum numbers $\ell,m$. The functions $S_{\ell,a\omega}^{m}$ are 
eigenfunctions of a self-adjoint operator. 

For the radial functions $R_{\ell,\omega}^{m}$ in (\ref{kerrrepresentation}) a {\it one-dimensional 
Schr\"odinger equation} of the form 
\begin{equation}\label{schro}
\frac{d^2{R}_{\ell,\omega}^{m}}{dr_{*}^2}+\left[\left(\omega-\frac{m a}{r^2+a^2}\right)^2
+\left(r-r_H\right)\cdot V_{\ell,\omega}^{m}(r_{*})\right]\,{R}_{\ell,\omega}^{m}=0\,,  
\end{equation}
can be derived from  (\ref{KG}), with suitable real potentials  $V_{\ell,\omega}^{m}(r_{*})$.

The ``physical solutions'' to (\ref{schro}) are supposed to possess the asymptotic behavior 
\begin{equation}\label{asympt}
{R}_{\ell,\omega}^{m}\sim 
\cases{
e^{-i\omega r_{*}} + \mathcal{R} \,e^{+i\omega r_{*}}  \;\;\mathrm{ as }\;\;r\rightarrow \infty
\cr
\mathcal{T}\,e^{-i(\omega- m \Omega_{H})r_{*}}         \;\;\;\;\;\mathrm{ as }\;\;r\rightarrow r_H\,,
}  
\end{equation} 
where $\Omega_{H}$ stands for the angular velocity of the black hole with respect to the asymptotically 
stationary observers, while $\mathcal{R}$ and $\mathcal{T}$ denotes the reflection and transmission 
coefficients. Notice that these boundary conditions presumes the existence of a transmitted wave submerging 
into the ergoregion.

By evaluating the Wronskian of the associated fundamental solutions, ``close'' to infinity and ``close'' 
to the horizon, it can be shown that 
\begin{equation}\label{range0}
(\omega- m \Omega_{H})\,|\mathcal{T}|^2 =(1-|\mathcal{R}|^2)\,\omega\,.
\end{equation}

In virtue of (\ref{range0}) it follows that whenever $|\mathcal{R}|>1$---or equivalently whenever $|\mathcal{T}|$ 
does not vanish and the inequality 
\begin{equation}\label{range}
0<\omega<m\,\Omega_{H}
\end{equation}
\vskip-0.1cm{
holds---positive energy is supposed to be acquired by the backscattered scalar mode due to its interaction with 
the Kerr black hole.}

\section{Numerical studies of superradiance} 

So far our considerations have been restricted to the study of individual modes. However the investigation of the 
linear stability problem \cite{wald2,kaywald,LB,DR} taught us the lesson that statements which are valid at 
the level of individual modes typically do not imply statements for finite energy solutions composed by infinitely 
many modes. 

This section is to reveal some of our pertinent numerical results. Before proceeding let us mention that 
the first time domain studies of superradiance was carried out long time ago \cite{krivan, laguna}. The 
scale of energy extraction was found to be smaller than the estimates recalled above. The numerical results 
reported below were derived by making use of our code called GridRipper which is a fully spectral in the 
{angular} directions while the dynamics in the complementary 1+1 Lorentzian spacetime is followed by making 
use of a fourth order finite differencing scheme \cite{gridripper3,cspalir} (see also \cite{csizmadia}).

\subsection{The initial data}

To have an incident scalar wave---to study the way a to be superradiant solution acquires extra energy by 
submerging into the ergoregion---, in a sufficiently small neighborhood of the initial data surface in 
the asymptotic region, the solution was assumed to posses the form
\begin{eqnarray}
\label{minkowskisolution}\Phi(t,r_{*},\vartheta,\tilde{\varphi}) 
\approx e^{-\I\,\omega_{0}\,(r_{*}-r_{*0}+t)} f(r_{*}-r_{*0}+t)\,Y{}_{\ell}^{m}(\vartheta,\tilde{\varphi}).  
\end{eqnarray}
where $f:\R\rightarrow\C$ is a smooth function of compact support and $\omega_{0}$, $r_{*0}$ are real parameters.
This suggest the use of initial data
\begin{eqnarray} 
&&\hskip-.5cm\phi(r_{*},\vartheta,\tilde{\varphi})=e^{-\I\,\omega_{0}\,(r_{*}-r_{*0})}
f(r_{*}-r_{*0})\,Y{}_{\ell}^{m}(\vartheta,\tilde{\varphi})\,,\nonumber\cr
&&\hskip-.65cm\phi_{t}(r_{*},\vartheta,\tilde{\varphi})=-\I\,\omega_{0}\,\phi(r_{*},\vartheta,\tilde{\varphi})
+e^{-\I\,\omega_{0}\,(r_{*}-r_{*0})}\,f'(r_{*}-r_{*0})\,Y{}_{\ell}^{m}(\vartheta,\tilde{\varphi})\,,\nonumber
\end{eqnarray} 
where $f'$ denotes the first derivative of $f:\R\rightarrow\C$. The Fourier transform, 
${\mycal{F}}\hskip-0.5mm\Phi$, of the approximate solution (\ref{minkowskisolution}) reads as 
\begin{equation}
{\mycal{F}}\hskip-0.5mm \Phi(\omega,r_{*},\vartheta,\tilde{\varphi}) 
\approx e^{-\I\,\omega\,(r_{*}-r_{*0})} {\mycal{F}}\hskip-1mm 
f(\omega-\omega_{0})\,Y{}_{\ell}^{m}(\vartheta,\tilde{\varphi}),  
\end{equation}
where $\omega$ is the temporal frequency and ${\mycal{F}}\hskip-1mm f$ stands for the Fourier-transform of $f$. 
Notice that ${\mycal{F}}\hskip-1mm f$ plays the role of a frequency profile, which guaranties that whenever 
${\mycal{F}}\hskip-1mm f$ is chosen to be a sufficiently narrow the approximate solution (\ref{minkowskisolution}) 
has to be close to a monochromatic wave packet, which for suitable value of $\omega_0$ becomes to be superradiant 
(for more details see \cite{cspalir}). 

\subsection{Numerical results}

The plots shown below refer to evolution of a pure quadrupole type initial data 
with radial profile function $f:\R\rightarrow\C$ 
\begin{eqnarray}
\label{profilefunction}
\hskip-0.3cm{f_{w}(x)=
\left\{
\begin{array} {r l}
e^{\left[-\left\vert\frac{w}{x+\frac{w}{2}}\right\vert-\left\vert\frac{w}{x-\frac{w}{2}}\right\vert+4\right]}\,, 
& \mathrm{if } x\in[-\frac{w}{2},\frac{w}{2}]\cr
 0\,, & \mathrm{otherwise} \,,
\end{array}
\right.}
\end{eqnarray}
which is a smooth function of the real variable $x$ with compact support $[-\frac{w}{2},\frac{w}{2}]$, 
and with initial parameters $M=1$, $a=0.99$,  $\ell=m=2$,\\  $\omega_{0}=\frac{1}{2}m\Omega_{H}$, 
$r_{*0}=31.823$. 

\begin{figure}
  \begin{center}
  \includegraphics[width=7.2cm]{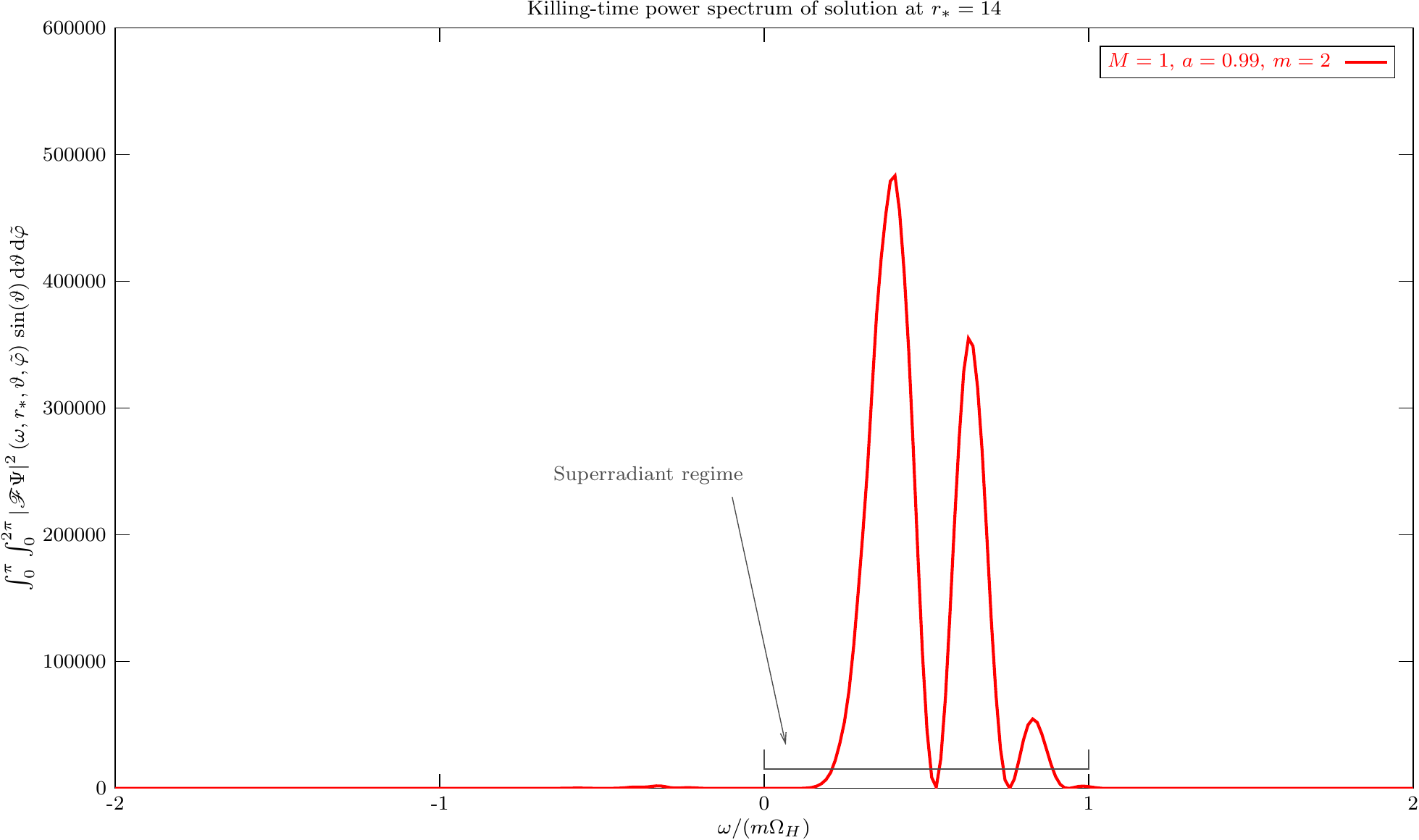}
  \end{center}
  \caption{\label{power} 
The power spectrum of the to be superradiant solution at $r_{*}=14$, located between the compact support and the black hole.}
\end{figure}

It is important to be sure (see Fig.\ref{power}) that the above choice yield a solution with the expected frequency profile. 

\begin{figure}
  \begin{center}
  \includegraphics[width=7.2cm]{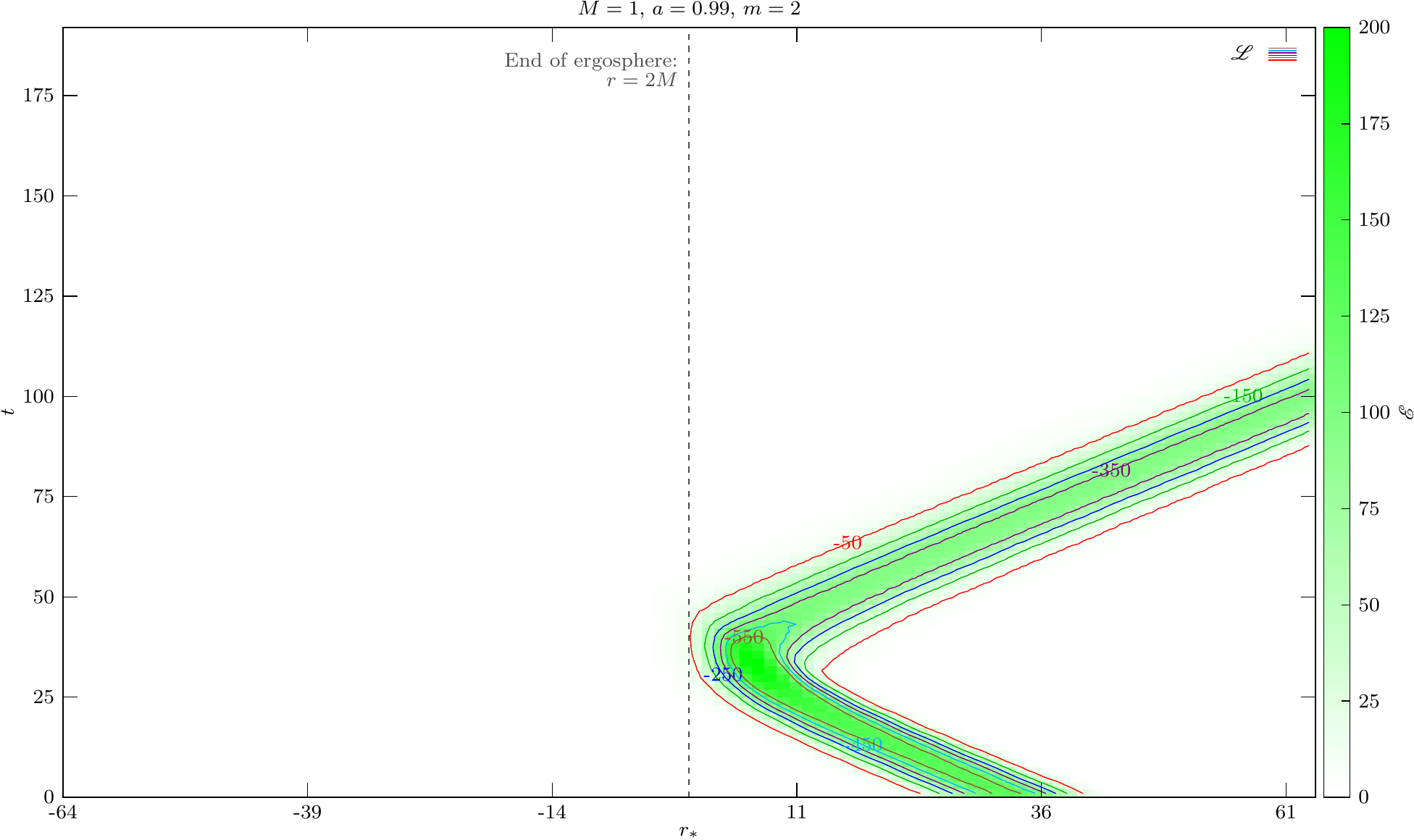} 
  \end{center} 
  \caption{\label{reflection} The radial energy and angular momentum  distributions.}
\end{figure}

\begin{figure}
  \begin{center}
  \includegraphics[width=7.2cm]{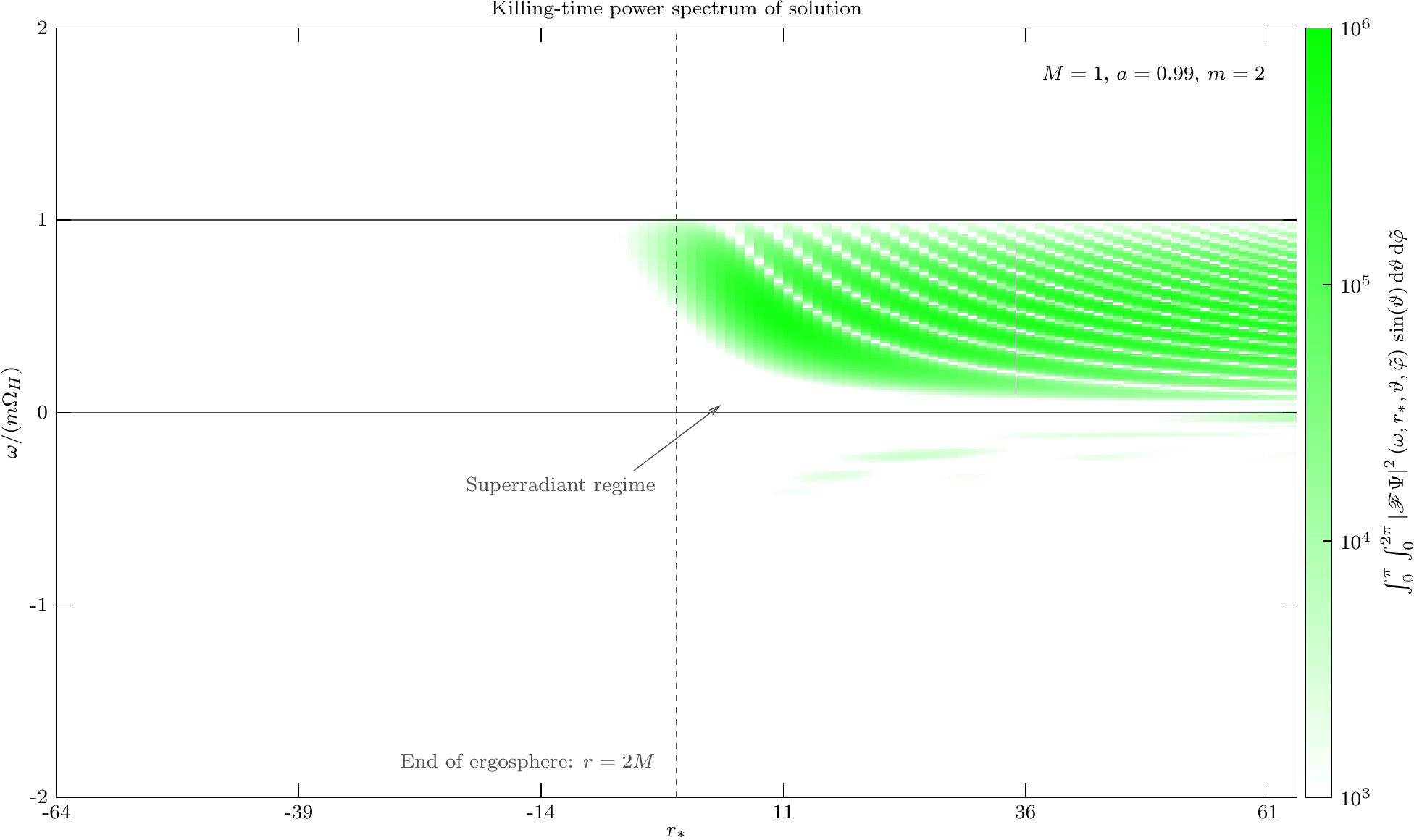} 
  \end{center} 
  \caption{\label{reflection_power} The radial distribution of the power spectrum.}
\end{figure}

The time dependence of the radial energy  and angular momentum  distributions, along with the complete power 
spectrum, are shown\footnote{Note that on all the included 2-dimensional plots the indicated quantities are 
integrated with respect to the radial degrees of freedom.} on the succeeding figures, Fig.\ref{reflection} and 
\ref{reflection_power}. These figures indicate that the reported nearly perfect reflection do really happen for the 
considered to be superradiant solution. 

\begin{figure}
  \begin{center}
  \includegraphics[width=7.2cm]{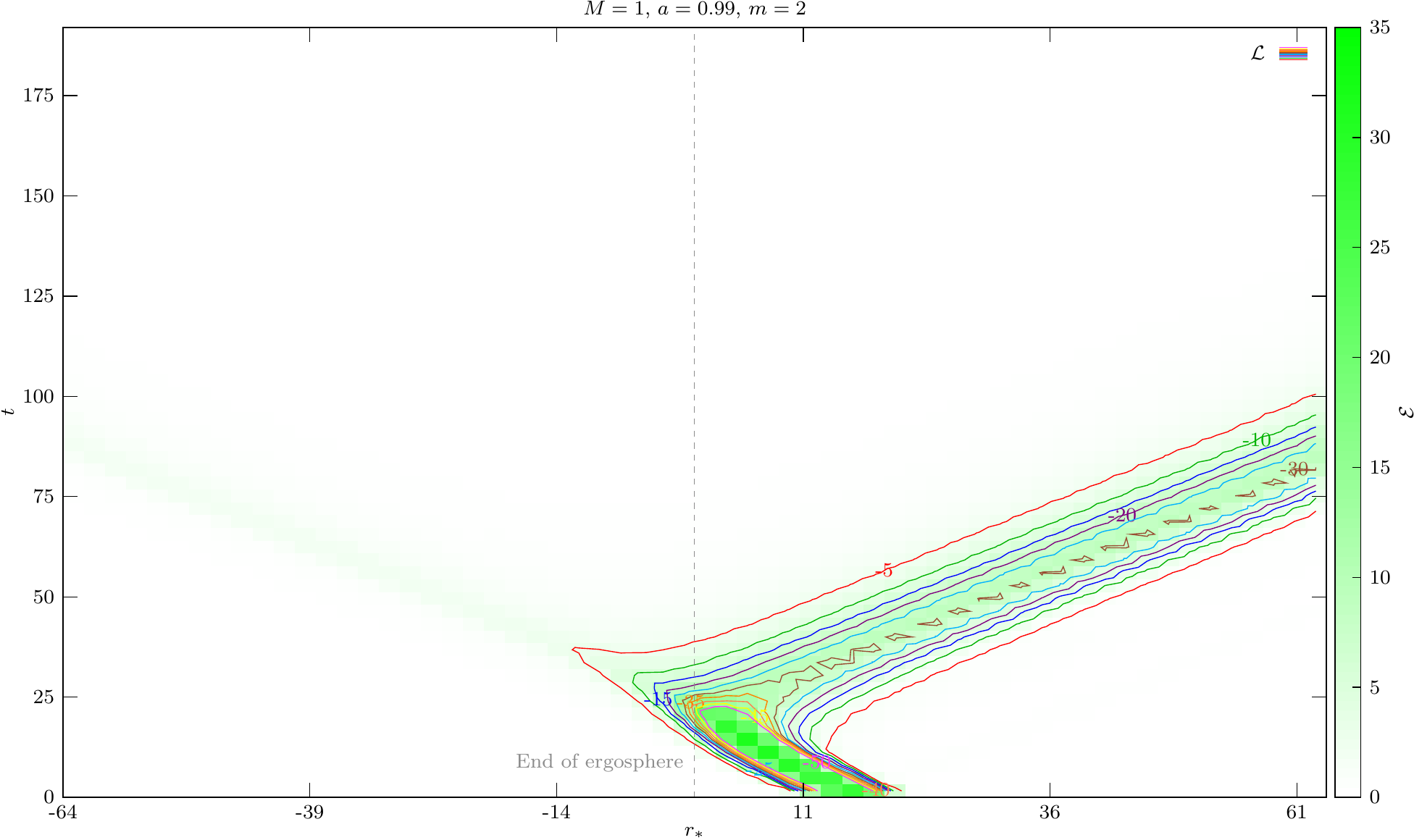}
  \end{center} 
  \caption{\label{reflection2}The frequency $\omega_{0}$ of this nearly superradiant solution is the same as before. Only the 
support is shifted to get a submerging part.}
\end{figure}

\begin{figure}
  \begin{center}
  \includegraphics[width=7.2cm]{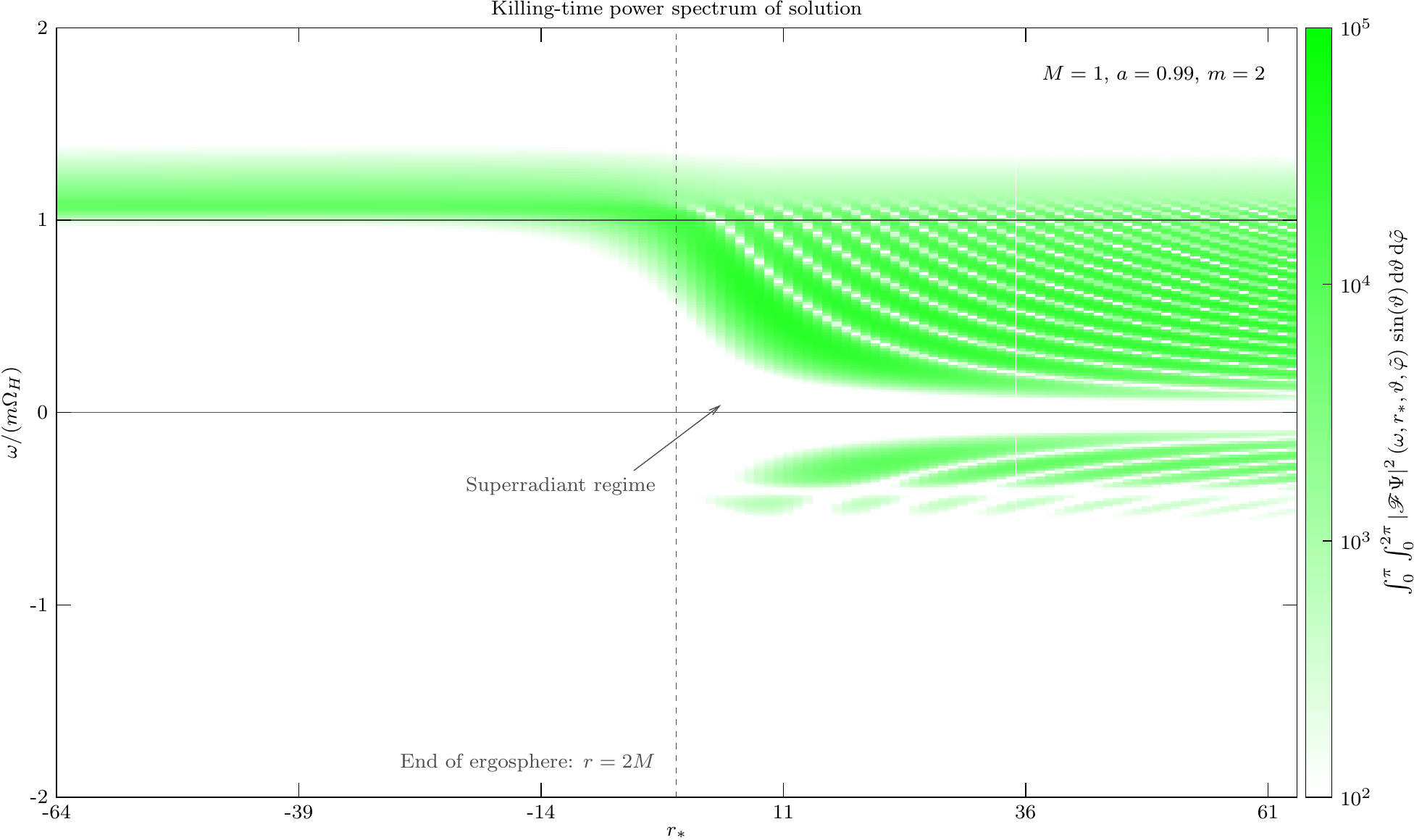}
  \end{center} 
  \caption{\label{reflection_power2} The untuned solution has a submerging part but 
its  power spectrum jumps out of the superradiant domain on reaching the ergoregion.}
\end{figure}

It is also informative to have a look at the corresponding figures (see Fig.\ref{reflection2} and \ref{reflection_power2}) 
for an almost to be superradiant solution yielded by shifting the compact support towards the black hole---decreasing slightly
thus the angular momentum of the radiation with respect to its energy---while all the other parameters remained intact. 
Notice that, in virtue of Fig.\ref{reflection_power2}, the frequency content of the part of the incident wave packet, 
submerging into the ergoregion, gets completely evacuated from the superradiant domain. 

It is also important to be mentioned that according to the accuracy of our code no energy extraction---or, at least, not 
more than $10^{-3}$ times of the initial energy---happened in either of these (or analogous) simulations.

\section{Summary} 
 
The numerical evolution of massless Klein-Gordon field on Kerr background, arising from initial data with compact support in 
the asymptotic region, was considered. The incident wave packet was tuned to maximize the effect of superradiance.

For perfectly tuned initial data no energy extraction could be observed. Significant part of the incident
radiation fail to reach the ergoregion and the time evolution mimics the phenomenon of a total reflection.

\begin{figure}
  \begin{center}
  \includegraphics[width=7.2cm]{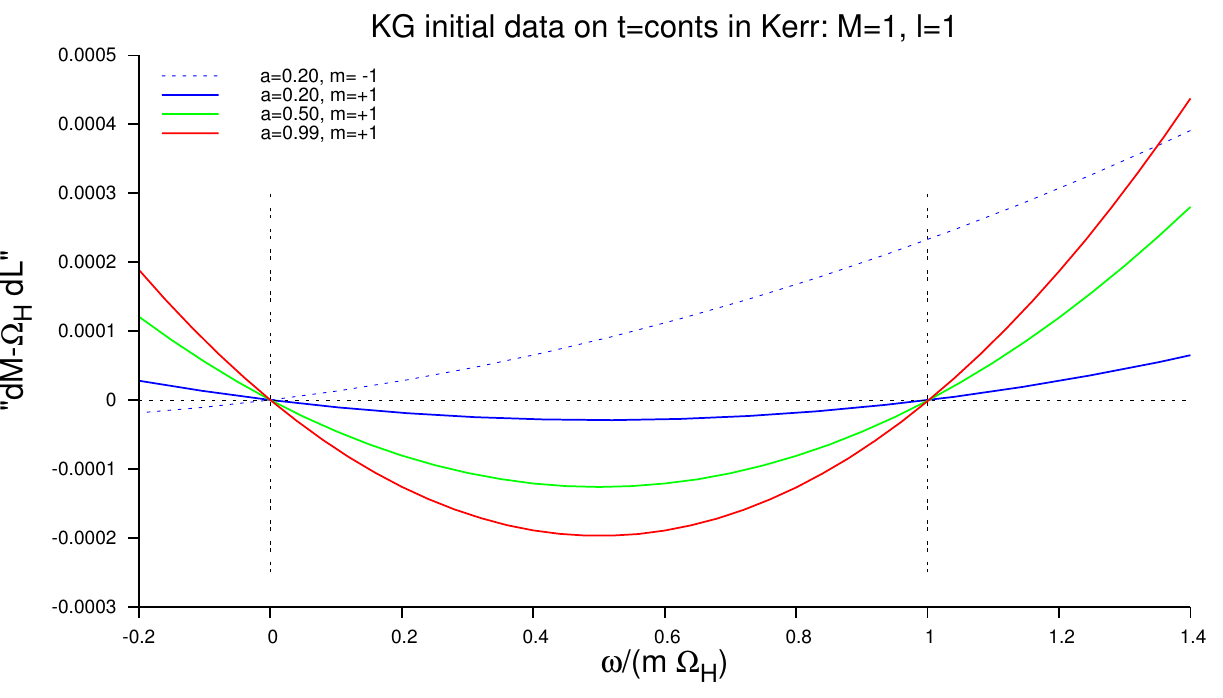}
  \end{center} 
  \caption{\label{EL_initialdata} The energy and angular momentum content of initial data specifications are compared. 
For to be superradiant configurations the incident wave cannot deliver its full energy and angular momentum to the black 
hole unless the second law of black hole thermodynamics is violated.}
\end{figure}

To get some insight about the physical mechanism beyond the nearly total reflection it turned out to be useful to 
compare the energy and angular momentum content of initial data specifications. Fig.\ref{EL_initialdata} is to demonstrate 
that far too much angular momentum is stored by the to be superradiant wave packets, as $dE< \Omega_H\,dL$ holds for them, which---in 
virtue of the second law of black hole thermodynamics---does not allow these wave packets to enter the black hole region. 
Accordingly, the observed nearly total reflection may be considered as the field theoretical analog of the phenomenon 
in Wald's thought experiments \cite{wald3} demonstrating, in the early 70', that a Kerr black hole does not capture a 
particle that would cause a violation of the relation $m^2\geq a^2+e^2$.

\section*{Acknowledgments} 
This research was supported in part by OTKA grant K67942.

\section*{References}
\bibliography{racz_prague}

\end{document}